\documentclass[10pt,twocolumn,twoside]{IEEEtran}
\IEEEoverridecommandlockouts
\usepackage{cite}
\usepackage{amsmath,amssymb,amsfonts}
\usepackage{algorithmic}
\usepackage{graphicx}
\usepackage{textcomp}
\usepackage{xcolor}
\usepackage{algorithm}
\usepackage{comment}
\usepackage{setspace}
\usepackage{color}
\usepackage{subfigure}
\newtheorem{theorem}{\bf Theorem}
\newcommand{\bb}{\boldsymbol{b}}
\newcommand{\bx}{\boldsymbol{x}}
\newcommand{\bz}{\boldsymbol{z}}

\newcommand{\bmu}{\boldsymbol{\mu}}
\newcommand{\bLambda}{\boldsymbol{\Lambda}}
\newcommand{\bPi}{\boldsymbol{\Pi}}

\newcommand{\cG}{{\cal G}}

\def\BibTeX{{\rm B\kern-.05em{\sc i\kern-.025em b}\kern-.08em
    T\kern-.1667em\lower.7ex\hbox{E}\kern-.125emX}}
\begin{document}
\title{Distributed Resource Allocation for Network Slicing of Bandwidth and Computational Resource}

\author{\IEEEauthorblockN{Anqi~Huang\IEEEauthorrefmark{1}, Yingyu~Li\IEEEauthorrefmark{1}, Yong~Xiao\IEEEauthorrefmark{1}, Xiaohu Ge\IEEEauthorrefmark{1}, Sumei Sun\IEEEauthorrefmark{2}, Han-Chieh Chao\IEEEauthorrefmark{3} \\
\IEEEauthorblockA{\IEEEauthorrefmark{1}School of Electronic Information and Communications, Huazhong Univ. of Science \& Technology, China}\\
\IEEEauthorblockA{\IEEEauthorrefmark{2} Institute for Infocomm Research, Singapore}\\
\IEEEauthorblockA{\IEEEauthorrefmark{3} School of Electrical Engineering, National Dong Hwa University, Taiwan}
}}
\maketitle

\begin{abstract}
Network slicing has been considered as one of the key enablers for 5G to support diversified services and application scenarios. This paper studies the distributed network slicing utilizing both the spectrum resource offered by communication network and computational resources of a coexisting fog computing network. We propose a novel distributed framework based on a new control plane entity, regional orchestrator (RO), which can be deployed between base stations (BSs) and fog nodes to coordinate and control their bandwidth and computational resources. We propose a distributed resource allocation algorithm based on Alternating Direction Method of Multipliers with Partial Variable Splitting (DistADMM-PVS). We prove that the proposed algorithm can minimize the average latency of the entire network and at the same time guarantee satisfactory latency performance for every supported type of service. Simulation results show that the proposed algorithm converges much faster than some other existing algorithms. The joint network slicing with both bandwidth and computational resources can offer around 15\% overall latency reduction compared to network slicing with only a single resource.
\end{abstract}

\vspace{-0.202in}
\section{Introduction}
\label{INTRODUCTION}
It is commonly believed that 5G will be much more than a simple upgrade of physical performance metrics such as throughput and capacity \cite{NGMN5GWhitePaper} \cite{ge20165g}. It will represent a fundamental transformation from the traditional data-oriented architecture towards a more flexible and service-oriented architecture. The {\em service-based architecture} (SBA) has been introduced by 3GPP as the key enabler for supporting a plethora of different services with diverse requirements on a common set of physical network resources. The core idea is to use software-defined networking (SDN) and network functions virtualization (NFV) to virtualize the network elements into network functions \cite{ETSINFV2017}, each of which consists of a functional building block utilizing various resources offered by the network. Each type of services can then be instantiated by a series of network function sets, called {\em network slice} \cite{NGMN2016NetworkSlicing}. Network slicing has been considered as the foundation of 5G SBA to match with diversified service requirements and application scenarios.

To support emerging computationally intensive applications, create new business opportunities and increase revenues, fog computing has recently been promoted by both industry and standardization institutions as one of the key components in 5G \cite{Chiang2017FogBook}. Compared to massive-scale cloud data centers that are typically built in remote areas, fog computing consists of a large number of small computing servers, commonly referred to as fog nodes, that can offload computationally intensive tasks closer to end user equipments (UEs) \cite{ge2018joint}. Fog computing networks can be deployed by cloud service providers such as Amazon and Microsoft. It can also be implemented by mobile network operators (MNOs) within their network infrastructure.

Recently, network slicing utilizing both communication and computational resources has attracted significant interest. Allowing each slice to be supported by both resources can further improve the overall UE experience, balance resource utilization across different network elements, and open doorways for newly emerging services with stringent latency and computational requirements \cite{zhong2017heterogeneous} \cite{ge2015spatial}. In spite of its great promise, allocating resources for multiple network slices with different resources introduces many new challenges. First, different resources are typically managed by different service providers. Therefore, exchanging and sharing proprietary information such as resource availability and traffic dynamics between them are generally impossible. Second, both fog computing and communication network infrastructure can be distributed in a wide geographical area and centralized coordination and management may result in intolerable coordination delay and excessive communication overhead. Finally, each UE can simultaneously request multiple types of services with different features offered by different resources. How to design an optimal algorithm that can quickly and accurately allocate various combination of different resources \cite{caballero2019network} to support multiple network slices remains an open problem.

In this paper, we investigate the distributed network slicing for a 5G system consisting of a set of base stations (BSs) offering wireless communication services and a coexisting fog computing network performing computationally-intensive tasks. We consider joint resource allocation of both bandwidth of BSs and processing power of fog nodes for supporting multiple network slices. We focus on reducing the overall latency experienced by end UEs that include both communication delay in wireless links connecting UEs and BSs and queuing delay at fog nodes. A distributed resource allocation algorithm has been proposed. We prove that the proposed algorithm can minimize the average latency of the entire network and at the same time guarantee satisfactory performance for each supported type of service. 
%
%
The main contributions of this paper are summarized as follows:
\begin{itemize}
\item[1)] A novel distributed framework based on regional orchestrator (RO) has been proposed for supporting distributed network slicing in a large network.

\item[2)] A distributed optimization algorithm based on Alternating Direction Method of Multipliers with Partial Variable Splitting (DistADMM-PVS) has been proposed to distributedly coordinate the resource allocation of both bandwidth of BSs and computational resources of fog nodes without requiring exchanging any proprietary information between BSs and fog nodes. We prove that the proposed algorithm can converge to the global optimal solution at a rate of $O(1/t)$.

\item[3)] Simulation and detailed performance analysis have been presented under various practical scenarios. Our result shows that joint slicing utilizing both bandwidth and computational resources offers around 15\% overall latency reduction compared to network slicing with only a single resource.
\end{itemize}

The rest of the paper is organized as follows. Related works are reviewed in Section \ref{Section_RelatedWork}. In Section \ref{Section_Model}, we present the system model. The network slicing architecture and RO are introduced in Section \ref{Section_Architecture}. Distributed optimization algorithm is presented in Section \ref{Section_Algorithm}. We present the simulation result in Section \ref{Section_SIMULATION} and conclude the paper in Section \ref{Section_Conclusion}.
\begin{spacing}{-0.82}
\end{spacing}
\section{Related Work}
\label{Section_RelatedWork}
One of the main challenges for network slicing is how to quickly and effectively isolate and distribute resources according to the specific requirement of each service. In \cite{jiang2016network}, the authors studied the allocation of the radio resources for network slicing. A prioritized admission mechanism was proposed to improve the resource utilization and increase UE's service experience. Network slicing has been studied for dynamic resource demand and availability in mobile environment in \cite{zhang2017network}. In \cite{xiao2018distributed}, the authors proposed a network slicing architecture utilizing spectrum resources in both licensed and unlicensed bands.

Recently, network slicing utilizing computational resource has attracted significant interest \cite{zhang2017computing}\cite{8437204}. In \cite{zhang2017computing}, the authors proposed a distributed joint optimization algorithm for the allocation of fog computing resources and applied it to improve the performance of Internet-of-Things (IoT) system. In \cite{8437204}, the authors proposed a computational resource allocation scheme based on double-matching for fog computing networks.
\begin{spacing}{-0.82}
\end{spacing}
\section{System Model}
\label{Section_Model}
\begin{figure}
  \centering
  \includegraphics[width=7.3cm]{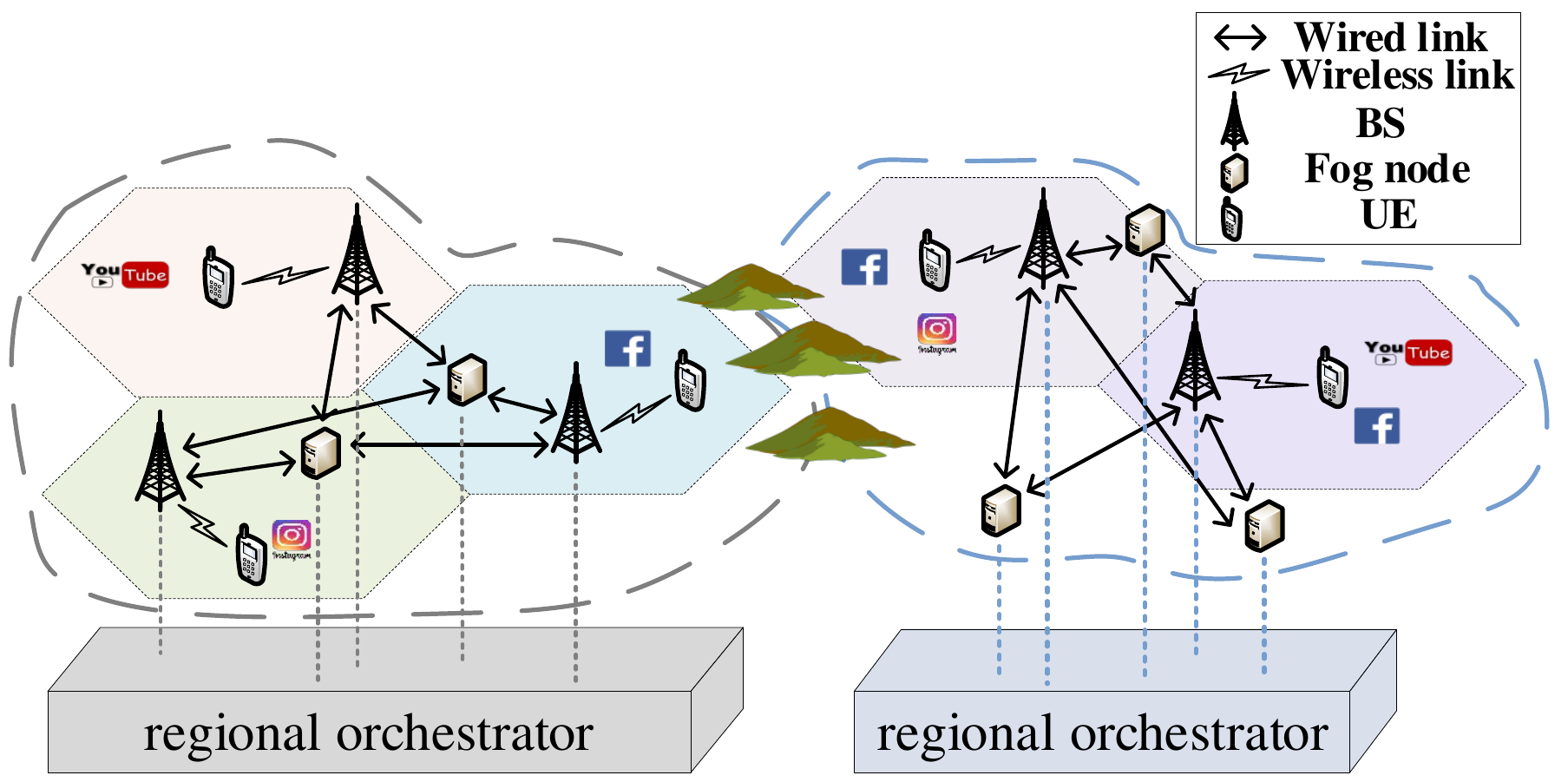}
  \vspace{-0.1in}
  \caption{Distributed network slicing model}
  \begin{spacing}{-0.6}
\end{spacing}
  \label{fig1}
\end{figure}

We consider a network system consisting of a set $ \mathcal{F}=\{1\cdots F\}$ of $F$ fog nodes and a set $\mathcal{S}=\{1\cdots S\}$ of $S$ BSs as illustrated in Figure 1. In a cellular network, each BS offers services in an exclusive coverage area. Suppose each UE can request at most $N$ types of services. Let $\mathcal{N}=\{1\cdots N\}$ be the set of all supported types of services. For each type of service, we assume that there exists a minimum volume of data, called task unit that can be transmitted by BSs and processed by fog nodes. For example, in video or audio processing service, each video or audio clip consists of a number of video or audio data units for transporting and processing. Let $d_n$ be the data size of each task unit of service type $n$. Each BS $s$ has been allocated with a fixed bandwidth, labeled as $\beta_s$ for all $s\in \mathcal{S}$, and each fog node $f$ can process at most $\mu_f$ task units per second for all $f\in \mathcal{F}$. Suppose the arrival rate $k_{sn}$ of the $n$th service task unit at BS $s$ follows a Poisson distribution, $k_{sn}\sim P(\lambda_{sn})$, where $\lambda_{sn}$ is the expected number of received task units.

In this paper, we consider joint resource allocation for multiple network slices. Network slicing employs a network virtulization approach that virtualizes physical resources into Virtual Network Functions (VNFs). Each VNF can be further divided into smaller components to be placed in a common software container, so the network functionality can be quickly released and reused by different service instances. We refer to the smallest component that can be used in the VNFs for network slices as a slice unit. Each network slice can consist of many slice units. Different slice units are decoupled from each other. So each network slice can be launched and dynamically scaled without affecting other ongoing services.

In this paper, we focus on minimizing the service response time for each type of service, which can consist of both communication delay for task unit transportation from UEs to fog nodes as well as the queuing delay at fog nodes. Let us first consider the communication delay. Note that in many practical networks, BSs and fog nodes can be connected with wireline or fiber which typically offers much higher data rate than the wireless links between UEs and BSs. Therefore, in this paper, we follow a commonly adopted setting and ignore the communication delay between BSs and fog nodes \cite{7239522}. For a given bandwidth $0\leq b_{sn}<\beta_s$ allocated by BS $s$ for service type $n$, follow the commonly adopted setting \cite{6773024}, the communication delay for transporting each unit of task can be written as
\begin{eqnarray}
p_{sn}={d_{n} \over {b_{sn} \cdot \log(1+h_{sn}\frac{w_{sn}}{\sigma_{sn}})}},
\label{communication delay}
\end{eqnarray}
where $w_{sn}$ is the required transmission power to transmit the task units for service type $n$ from the UEs to  BS $s$, $\sigma_{sn}$ is the additive noise level received at BS $s$, and $h_{sn}$ is the channel gain between BS $s$ and the associated UE for service type $n$.

Queuing delay at the fog node can be affected by the processing power of fog nodes and task arrival rate. Suppose the maximum processing power allocated by fog nodes to process the $n$th type of service offered by BS $s$ for its associated UEs is $\mu_{sn}$. We follow a commonly adopted setting and assume the task units processed by fog nodes can be modeled as M/M/1 queuing \cite{edelson1975congestion}. We can write the queuing delay of the $n$th type of service offered to UEs in the coverage area of BS $s$ as
\begin{eqnarray}
q_{sn}={1 \over {\mu_{sn}-\lambda_{sn}}}.\label{queuing delay}
\end{eqnarray}

By combining (\ref{communication delay}) and (\ref{queuing delay}), the overall service response time of the $n$th type of service offered by BS $s$ can be written as
\begin{eqnarray}
t_{sn}=p_{sn}+q_{sn}.
\end{eqnarray}
\begin{spacing}{-1.15}
\end{spacing}
\section{Distributed Network Slicing Architecture}
\label{Section_Architecture}
As mentioned earlier, two of the main challenges for quickly and accurately allocating resources across both communication and computing networks are:
\begin{itemize}
\item[1)] Physical resources can be arbitrarily deployed over a large geographical area and therefore a centralized resource management and control architecture may result in intolerably high latency and excessive communication overhead.

\item[2)] Communication network infrastructure and fog computing networks can be owned by different service providers. Therefore, proprietory information cannot be shared between these two systems.
\end{itemize}

The above challenges cannot be addressed by simply extending the existing centralized SDN control plane frameworks such as OpenFlow\cite{ONFSDNArchitecture} into a distributed setting. Actually, it has been observed by many existing works that OpenFlow-based SDN controller has been designed to mainly focus on managing the routing of data traffics and establishing and maintaining interconnections between virtual mesh networks \cite{Tootoonchian2010SDNHyperflow}. It can be used to maintain network connection and service continuity even in mobile environment. It however cannot be applied to manage the computational resources of fog computing networks. Also, OpenFlow relies on a centralized SDN controller to manage network resources and can only establish static paths to each SDN switch.

In this paper, we proposed a distributed network slicing architecture based on a new control plane entity, RO, deployed between communication network and fog computing network to support the fine-grained control of resources across both network systems. In this architecture, the total coverage area has been divided into a set of sub-regions, each of which consists of a limited number of closely located BSs and fog nodes that can be connected with high-speed local wireline links. A RO can then be deployed in each sub-region to control a set of VNFs composed of locally available communication and computational resource units. Each RO can only control and instantiate network slicing with local VNFs within each sub-region.  
Each BS will query the RO with the resource requests whenever it receives a service task request from UEs. The RO will then coordinate with the service requesting BS and neighboring fog nodes to create the corresponding network slices. The RO will also supervise the path reservation and routing of the service traffic between BSs and fog nodes. Two or more ROs can coordinate with each other and jointly adjust the volume and distribution of their local VNFs if two or more neighboring sub-regions experience unbalanced traffic loads. Our proposed architecture is illustrated in Figure \ref{fig1}.

In 3GPP's network slicing framework, a certain amount of resource must be reserved and isolated for each supported type of service, so there always exist available resources whenever a service request has been received. In this paper, we consider the RO implemented in 3GPP's framework \cite{3GPPNetworkSlicing} \cite{3GPPManagement}. For a limited time duration, each RO must first reserve a certain amount of computational resources at fog nodes, which is given by $\lambda_{sn}$, and a limited bandwidth, which is given by $b_0$ for task units of service type n at BS $s$. During this time, the reserved resources will be utilized to support the requested service instances, so we have the following constraints:
\begin{subequations}
\begin{align}
&{b_{sn}} > {b_0},\\
&{\mu _{sn}} > {\lambda _{sn}}.
\end{align}
\vspace{-0.2in}
\end{subequations}

In this paper, we focus on the resource allocation and network slicing within a specific time duration in which the maximum bandwidth of a set of local BSs and a given amount of processing power of local fog nodes have been reserved for a set of supported types of services. The dynamic resources allocation and network slicing will be left for our future research.
We consider the following constraints:
\begin{itemize}
\item[1)] {\em Bandwidth constraint}: Let $\beta_s$ be the total bandwidth reserved by each BS $s$. In other words, the total bandwidth that can be allocated by BS $s$ to all upcoming service tasks cannot exceed $\beta_s$. Generally speaking, the RO needs to reserve sufficient resources without knowing the exact number of task units which will be arrived in the future. It is however possible for the RO to estimate the possible number of arrival task units according to the empirical probability distribution of the task arrival rate. In this way, the RO can reserve sufficient resource to support the performance-guaranteed services for the majority of possible tasks with a certain level of confidence. More specifically, we define the confidence level $\theta=\Pr(k_{sn}\leq \theta_{sn})$ as the possibility that the number of type $n$ service task units arrived at BS $s$ is below a certain threshold number $\theta_{sn}$. For example, $\theta=0.9$ means that the RO wants to reserve resources to meet the demands of all UEs with 90\% confidence. We can observe that $\theta$ is equivalent to the Cumulative Distribution Function (CDF) of task arrival rate $k_{sn}$. We can therefore write $\theta_{sn}=CDF_k^{-1}(\theta,\lambda_{sn})$ where $(\cdot)^{-1}$ is the inverse function. 
    We can then have the following constraint for the bandwidth allocated by BS $s$ to a set $\mathcal{N}$ of all supported types of services
\begin{eqnarray}
\sum _{n\in \mathcal{N}} \theta_{sn}\cdot b_{sn}\leq \beta_s.
\end{eqnarray}
\item[2)] {\em Computational resource constraint}: Suppose the total computational resource $\gamma$ reserved to all fog nodes in a sub-region is limited. The sum of the computational resources allocated to all the services cannot exceed $\gamma$. We then have the following computational resources constraint
\begin{eqnarray}
\sum _{s\in \mathcal{S}}\sum _{n\in \mathcal{N}}\mu_{sn}\leq \gamma.
\end{eqnarray}
\end{itemize}

In addition, we assume each supported type of service $n$ has a maximum tolerable latency, labeled as $\overline {{T_{n}}}$, i.e., we have ${t_{sn}} \le \overline {{T_{n}}},\forall {s\in \cal S},{n \in \cal N}$.

Our proposed architecture is general and flexible. It can be applied to network slicing utilizing multiple resources across a wide geographical area. In this paper, we consider network slicing architecture to leverage the RO.
BSs and fog nodes can coordinate with the RO to dynamically allocate the bandwidth and the computational resources. In summary, we focus on designing a distributed algorithm to optimize the following problem
\begin{subequations}
\vspace{-0.1in}
\label{joint_prob}
\begin{align}
\min_{\{b_{sn}\}\{\mu_{sn}\}}  &\sum _{s\in \mathcal{S}}\sum _{n\in \mathcal{N}}t_{sn}\\
\mbox{ s.t. } \quad
&{b_{sn}} > {b_0},\forall {s\in \cal S},{n \in \cal N},\\
&{\mu _{sn}} > {\lambda _{sn}},\forall {s\in \cal S},{n \in \cal N},\\
&{t_{sn}} \le \overline {{T_{n}}},\forall {s\in \cal S},{n \in \cal N},\\
&\sum\limits_{n \in {\cal N}} {{\theta_{sn}}}  \cdot {b_{sn}} \le \beta_s,\forall {s\in \cal S},{n \in \cal N},\\
&\sum\limits_{s \in {\cal S}} {\sum\limits_{n \in {\cal N}} {{\mu _{sn}}} }  \le \gamma,\forall {s\in \cal S},{n \in \cal N}.
\end{align}
\end{subequations}
\begin{spacing}{-1.15}
\end{spacing}
\section{Distributed Optimization for Joint Network Slicing}
\label{Section_Algorithm}
As mentioned before, in order to minimize the overall latency experienced by end UEs, we need to carefully decide the resources allocated to each network slice.
However, solving problem (\ref{joint_prob}) involves jointly deciding the proper amount of bandwidth and processing power for every type of services with global information such as the expected number of arrived task units and the computational capacity of every fog node which may result in intolerably high latency and communication overhead.

To address the above problems, we need to develop a  distributed optimization algorithm for solving the joint network slicing problem in  (\ref{joint_prob}) with the following design objectives:
\begin{enumerate}
  \item {\it Distributed Optimization with Coordination:} The proposed optimization algorithm should be able to separate the global optimization problem into a set of sub-problems, each of which can be solved by a BS according to its local information. The RO can then be used to coordinate the solution of these subproblems to achieve the globally optimal resource allocation solution.
  \item {\it Privacy Preservation:} BSs and fog nodes may not want to share their private information such as bandwidths, expected number of arrived task units, and computational capacities with each other. 
  \item {\it Fast Convergence:} BSs and fog nodes connected to each RO can change over the time. Thus, the algorithm that needs to quickly converge to the global optimal solution.
\end{enumerate}

We propose a distributed optimization algorithm based on Alternating Direction Method of Multipliers (ADMM). Compared to traditional convex optimization algorithms, ADMM is more suitable for solving inequality constrained optimization problems in a decentralized manner. Furthermore, the decomposition-coordination procedure of the ADMM makes it possible to protect the aforementioned private information of BSs and fog nodes. Unfortunately, normal ADMM approaches can only handle problems with two blocks of variables \cite{boyd2011distributed}. In order to solve problem (\ref{joint_prob}) with the above objectives 1)-3), we propose a distributed ADMM algorithm with Partial Variable Splitting referred to as DistADMM-PVS. In this algorithm, the Lagrangian dual problem of  (\ref{joint_prob}) will be divided to $S$ sub-problems, each of which can be solved by an individual BS using its local information. The RO will collect the intermediate results from BSs and send the coordination feedbacks.    

Let us first follow the same line as \cite{boyd2011distributed} and combine constraints in (\ref{joint_prob}) with the objective function by introducing a set of $S+1$ indicator functions. Specifically, for constraints (\ref{joint_prob}b)-(\ref{joint_prob}e) that can be separated across different BSs, we define $\cG_s = \{\bb_{s  }, \bmu_{s  } : {b_{sn}} > {b_0}, {\mu _{sn}} > {\lambda _{sn}}, {t_{sn}} \le \overline {{T_{n}}}, \sum_{n \in {\cal N}} {{\theta_{sn}}}\cdot {b_{sn}} \le \beta_s,\forall {s\in \cal S}, {n\in \cal N}\}$ as the feasible set corresponding to BS $s$ where $\bb_{s  } = \langle b_{sn} \rangle_{n \in \cal N}$ is the vector of bandwidth allocated by BS $s$ for each type of services and $\bmu_{s  } = \langle \mu_{sn} \rangle_{n \in \cal N}$ is the vector of processing power allocated for each type of services connected to BS $s$. Let $\bx_{s  } =\langle \bb_{s  }, \bmu_{s  } \rangle, \forall s\in \cal S$, we define $S$ indicator functions as
\begin{eqnarray}
{\bf I}_{\cG_s} \left(\bx_{s  } \right) = \left\{ {\begin{array}{*{20}{c}}
{0,} & \bx_{s  } \in \cG_s, \\
{+\infty,} & \bx_{s  } \notin \cG_s.
\end{array}} \right.\forall s\in \cal S
\end{eqnarray}

For constraint (\ref{joint_prob}f) that cannot be separated, we can also define an indicator function ${\bf I}_{\cG} \left(\bmu \right)$ as
 \begin{eqnarray}
{\bf I}_{\cG} \left(\bmu \right) = \left\{ {\begin{array}{*{20}{c}}
{0,} & \bmu \in \cG, \\
{+\infty,} & \bmu \notin \cG,
\end{array}} \right.
\end{eqnarray}
 where $\bmu = [\bmu_1, \bmu_2, \ldots, \bmu_S]$, and $\cG$ is the half space defined by $\cG = \{\bmu: \sum_{s \in {\cal S}} {\sum_{n \in {\cal N}} {{\mu _{sn}}} }  \le \gamma\} $.

By including the above indicator functions ${\bf I}_{\cG_s}$ and ${\bf I}_{\cG} $, the original joint network slicing problem (\ref{joint_prob}) with a set of inequality constraints can be converted to the following form without inequality constraints
\begin{subequations} \label{Sep_Prob}
  \begin{align}
    \min_{\{\bx_{s  }, \bz_{s  }\}} & \quad \sum_{s\in {\cal S}} \left\{ f(\bx_{s }) + {\bf I}_{\cG_s} \left(\bx_{s} \right) \right\} + \bf I_{\cG} \left(\bz \right)      \label{Sep_Obj}             \\
    \mbox{ s.t. }    & \quad \bx_{s}= \bz _{s} , \forall s \in {\cal S},
  \end{align}
  \label{Prox_J}
\end{subequations}
where $f(\bx_{s }) = \sum_{n\in \cal N} t_{sn}$, and $\bz= [\bz_1, \bz_2, \ldots, \bz_S]$ is the introduced auxiliary variable.

The augmented Lagrangian of problem (\ref{Sep_Prob}) is given by
\begin{eqnarray}\label{Lag}
  \cal L_{\rho} (\bx, \bz, \bLambda) &=& \sum_{s\in {\cal S}} \left\{ f(\bx_{s }) + {\bf I}_{\cG_s} \left(\bx_{s} \right) \right\} + \bf I_{\cG} \left(\bz \right) \nonumber\\
   &+& \bLambda^T(\bx-\bz) + \frac{\rho}{2} ||\bx-\bz||_2^2 ,
\end{eqnarray}
where $\bLambda$ is the dual variable and $\rho$ is the augmented Lagrangian parameter.

We can then prove the following result.
\begin{theorem}
\label{Theorem_Convex}
The augmented Lagrangian of problem (\ref{Sep_Prob}) specified in (\ref{Lag})  is convex and partially separable among $\bx_s$.
\end{theorem}
\begin{IEEEproof}
For the convexity, it can be directly shown that set $\cG_s, \forall s \in \cal S$ and halfspace $\cG$ are all convex sets, and their intersection which is the feasible set of problem (\ref{Sep_Prob}) is also a convex set. We can also show that within the feasible set of problem (\ref{Sep_Prob}), the second derivative of $f(\bx_{s })$ is always positive which means that it is convex. Due to the fact that summation preserves convexity, we can then prove that $\cal L_{\rho} (\bx, \bz, \bLambda)$ is convex.

To prove (\ref{Lag}) is partially separable, we rewrite the augmented Lagrangian in (\ref{Lag}) as follows
\begin{eqnarray}
 {\cal L}_{\rho} (\bx_1, ..., \bx_S, \bz, \bLambda)= \sum_{s\in {\cal S}} \left\{ f(\bx_{s }) + {\bf I}_{\cG_s} \left(\bx_{s} \right) \right. \nonumber \\
  +\bLambda_s^T(\bx_s-\bz_s) + \frac{\rho}{2} ||\bx_s-\bz_s||_2^2 \left.\right\}  + \bf I_{\cG} \left(\bz \right),\label{Lp}
 \end{eqnarray}
From (\ref{Lp}), we can observe that ${\cal L}_{\rho} (\bx_1, ..., \bx_S, \bz, \bLambda)$ can be partially separated across $\bx_s$ for $s\in {\cal S}$.
 This concludes the proof.
\end{IEEEproof}

We can then convert problem (\ref{Sep_Prob}) into two-block ADMM form as follows
\begin{subequations}
\begin{align}
{\bx^{k + 1}} & = \mathrm{\mathop {argmin}\limits_{\bx}} \sum_{s\in {\cal S}} \left\{ f(\bx_{s }) + {\bf I}_{\cG_s} \left(\bx_{s} \right) \right\} \nonumber\\
&+ \frac{\rho }{2}\|\bx - \bz^{k}+\bLambda^{k}\|_2^2  \label{CADMM_x},\\
{\bz^{k + 1}} &= \mathrm{\mathop {argmin}\limits_{\bz}} \bf I_{\cG} \left(\bz \right)  + \frac{\rho }{2}\|\bx^{k+1} - \bz+\bLambda^{k}\|_2^2 \label{CADMM_z},\\
{\bLambda^{k + 1}} &=\bLambda^{k}+\bx^{k+1}-\bz^{k+1},\label{CADMM_lambda}
\end{align}
\end{subequations}
where $k$ denotes the number of iterations.  According to the partially separability of $\cal L_{\rho} (\bx, \bz, \bLambda) $,  we can divide (\ref{CADMM_x}) into a set of sub-problems, each of which can be solved by a BS using its local information. In particular, each BS $s$ solves the following sub-problem
\begin{eqnarray}
   \bx_s^{k + 1} &=& f(\bx_{s }) + {\bf I}_{\cG_s} \left(\bx_{s} \right) + \frac{\rho }{2}\|\bx_s - \bz_s^{k}+\bLambda_s^{k}\|_2^2, \nonumber\\
  && \forall s \in \cal S.\label{DistADMM_x}
\end{eqnarray}

Meanwhile, (\ref{CADMM_z}) is equivalent to projecting the point $\bx^{k+1}+\bLambda^{k}$ onto the halfspace $\cG$, i.e.
\begin{equation}\label{DistADMM_z}
  {\bz^{k + 1}} = {\bPi_{\cG}}(\bx^{k+1}+\bLambda^{k}),
\end{equation}
where $ {\bPi_{\cG}}(\cdot)$ denotes the projection onto halfspace $\cG$.

Detailed description of our proposed algorithm is presented in Algorithm 1.

\begin{theorem}
\label{Theorem_convergence}
The proposed DistADMM-PVS algorithm converges to the global optimal solution of the joint network slicing problem in (\ref{joint_prob}) at a rate of $O(1/t)$.
\end{theorem}
\begin{IEEEproof}
Since the distributed sub-problems specified in (\ref{DistADMM_x}) is equivalent to the centralized $\bx$-update in (\ref{CADMM_x}), the convergence property of our proposed DistADMM-PVS algorithm directly follows that of the standard ADMM approach \cite{boyd2011distributed}. Here we omit the details due to the limit of space.
\end{IEEEproof}

\begin{algorithm}
\footnotesize
  \caption{Distributed ADMM with Partial Variable Splitting (DistADMM-PVS)}\label{Algorithm 1}
  \begin{itemize}
    \item[] Initialization: Each BS $s$ chooses an initial variable $\bx_s^0$ and the RO chooses an initial dual variable $\bLambda^0$; $k=1$
    \item[] Set the maximum number of iterations as $K>0$
    \item[] {\bf while} $k \leq K$ {\bf do}
          \begin{itemize}
            \item[] 1. Each BS $s$ simultaneously do:
                  \begin{itemize}
                    \item[] 1) Update $\bx_s^{k+1}$ according to (\ref{DistADMM_x}) and report it to the RO;
                    \item[] 2) Allocate its bandwidth for each type of arrived task according to $\bb_s^{k+1}$;
                  \end{itemize}
            \item[] 2. After all the $\bx_s^{k+1}$ are received, the RO do:
                  \begin{itemize}
                    \item[] 1) Update auxiliary variable $\bz^{k+1}$ according to (\ref{DistADMM_z});
                    \item[] 2) Update dual variable $\bLambda^{k+1}$ according to (\ref{CADMM_lambda});
                    \item[] 3) {\bf if} {Stopping criteria met}
                    \item[]  \quad\quad break;
                    \item[] \quad {\bf end if}
                    \item[] 4) Sends the sub-vectors $\bz_s^{k+1}$ and $\bLambda_s^{k+1}$ to the corresponding BS $s$;
                  \end{itemize}
            \item[] 3. $k=k+1$
          \end{itemize}
    \item[] {\bf end while}
  \end{itemize}
\end{algorithm}
\vspace{-0.3in}
\section{Simulation Results And Comparative Analysis}
\label{Section_SIMULATION}
To evaluate the performance of our proposed network slicing architecture, we simulate a network system consisting of 285 BSs and 285 fog nodes deployed in Dublin which can support 3 types of services: text, audio, and video process service, as shown in Figure \ref{fig2a}. To prove the applicability of our proposed distributed network slicing architecture, we simulate 3 areas from city centers to suburbs as shown in Figure \ref{fig2b}, we assume each BS has reserved 30MHz bandwidth, and each fog node can process at most 180 task units per second. We assume BSs and fog nodes in the same area have reserved same bandwidth and computational resources. We compare the joint slicing involving both bandwidth of BSs and processing powers of fog nodes with two other network slicing scenarios, each of which only utilizes one type of resource, referred to as bandwidth slicing and computational resource slicing, respectively.

%
%

\begin{figure}
\begin{minipage}[t]{0.5\linewidth}
\centering
\includegraphics[width=4.5cm]{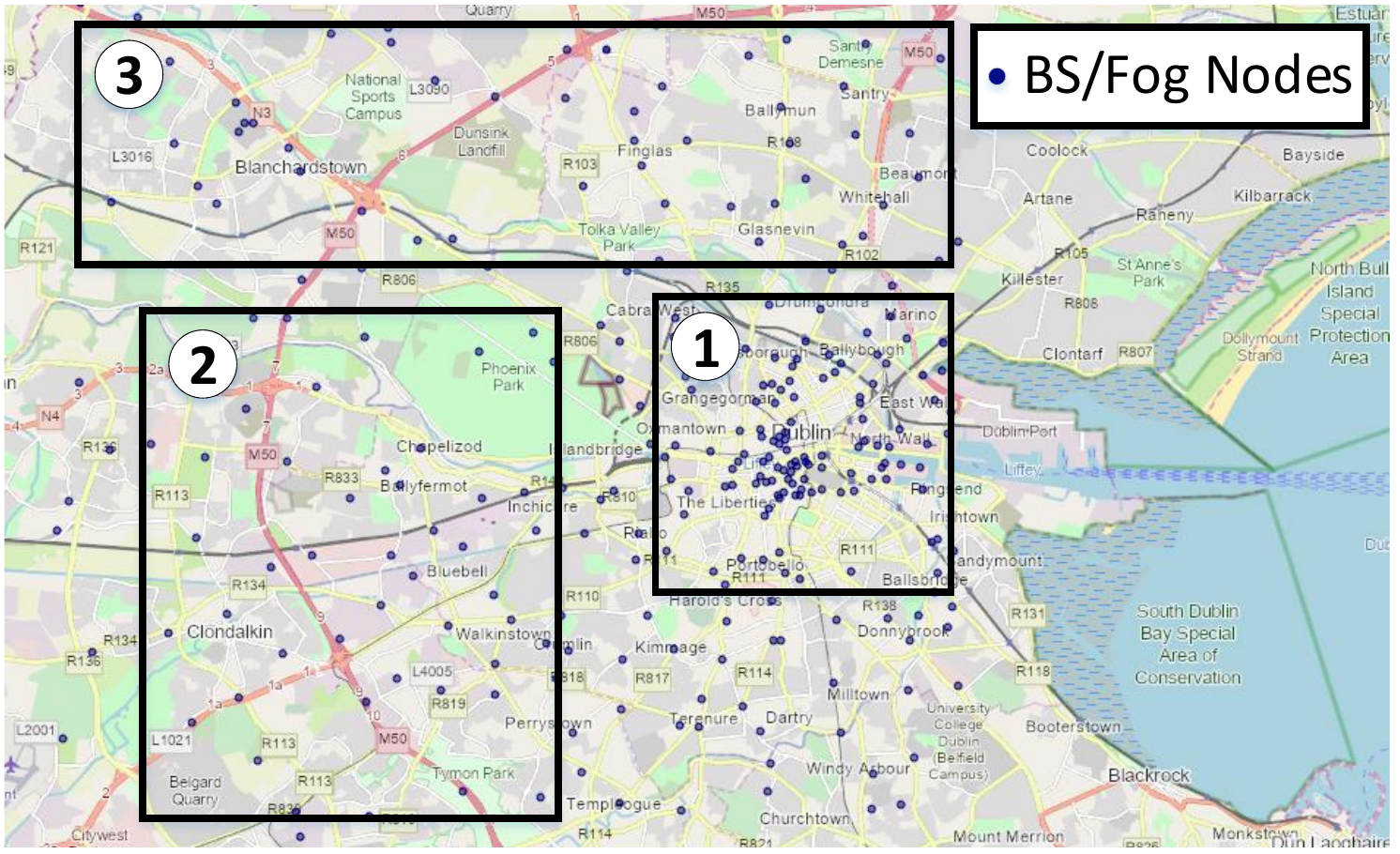}
\vspace{-0.3in}
\caption{Distribution of BSs}
\label{fig2a}
\end{minipage}
\begin{minipage}[t]{0.45\linewidth}

\centering
\includegraphics[width=4cm]{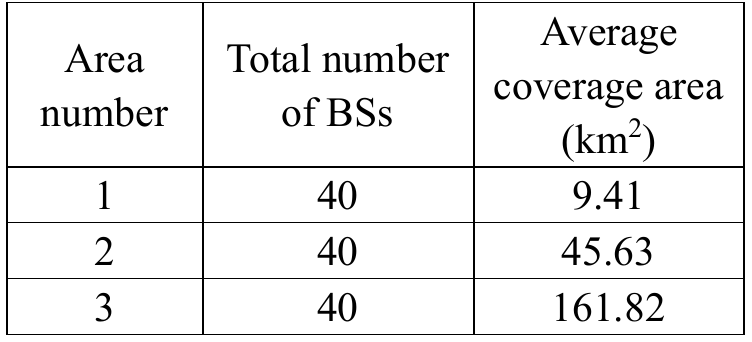}
\vspace{-0.3in}
\caption{Average Coverage Area of BSs}
\label{fig2b}
\end{minipage}

\vspace{-0.3in}
\end{figure}

\begin{figure*}[!htbp]
\setcounter{figure}{2}
 \subfigure{
 \begin{minipage}[t]{0.19\linewidth}
 \centering
 \includegraphics[width=3.8cm]{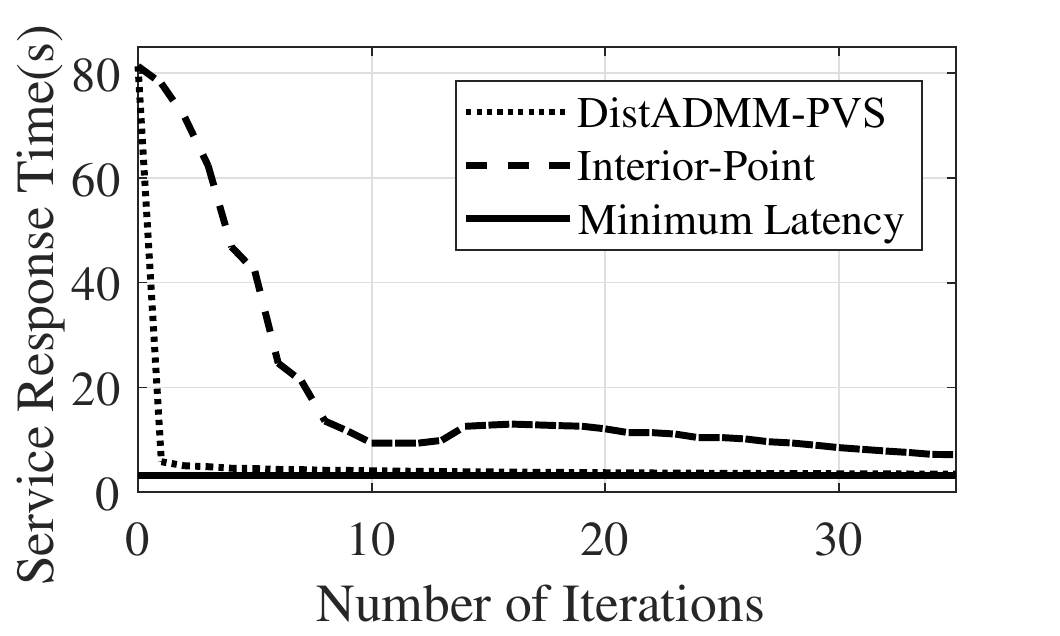}
 \vspace{-0.3in}
 \caption{Comparison of Interior-Point and Algorithm 1}
 \label{fig3}
 \end{minipage}%
 }%
 \subfigure{
 \begin{minipage}[t]{0.19\linewidth}
 \centering
 \includegraphics[width=3.8cm]{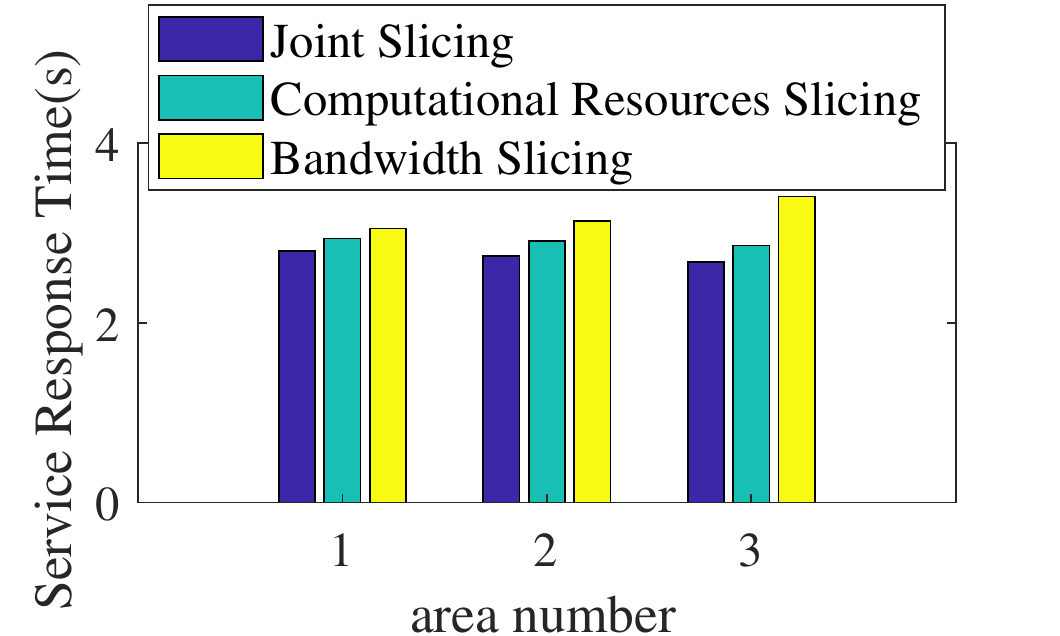}
 \vspace{-0.3in}
 \caption{Comparison of Three Architectures in Different Areas}
 \label{fig_area_comparison}
 \end{minipage}%

 }%
 \subfigure{
 \begin{minipage}[t]{0.19\linewidth}
 \centering
 \includegraphics[width=3.8cm]{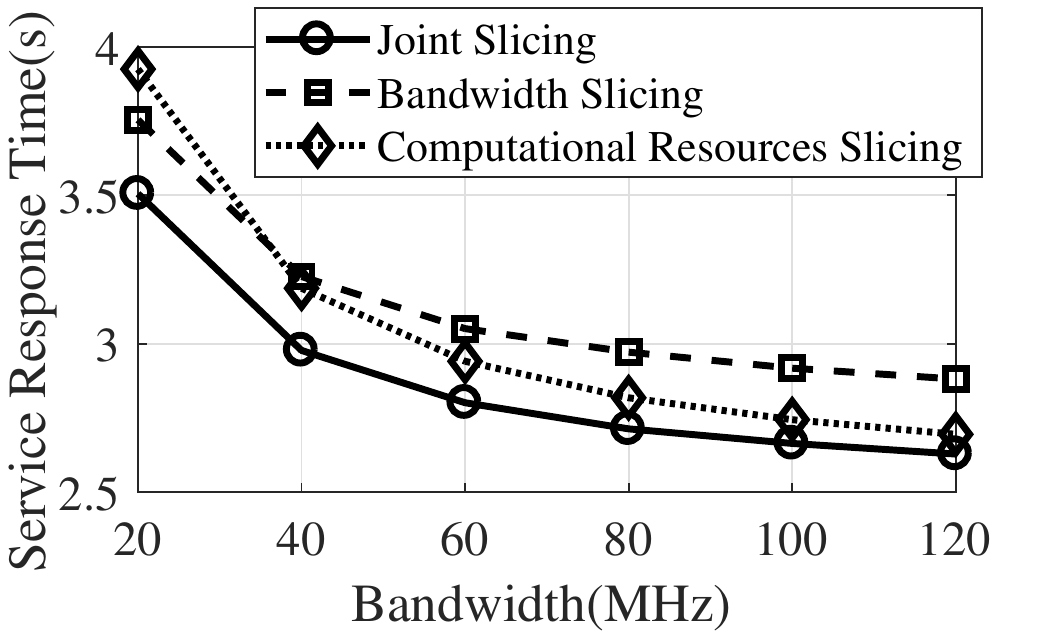}
 \vspace{-0.3in}
 \caption{Comparison of Three Architectures under Different Bandwidth of Each BS of Area1}
 \label{fig4}
 \end{minipage}%

 }
 \subfigure{
 \begin{minipage}[t]{0.19\linewidth}
 \centering
 \includegraphics[width=3.8cm]{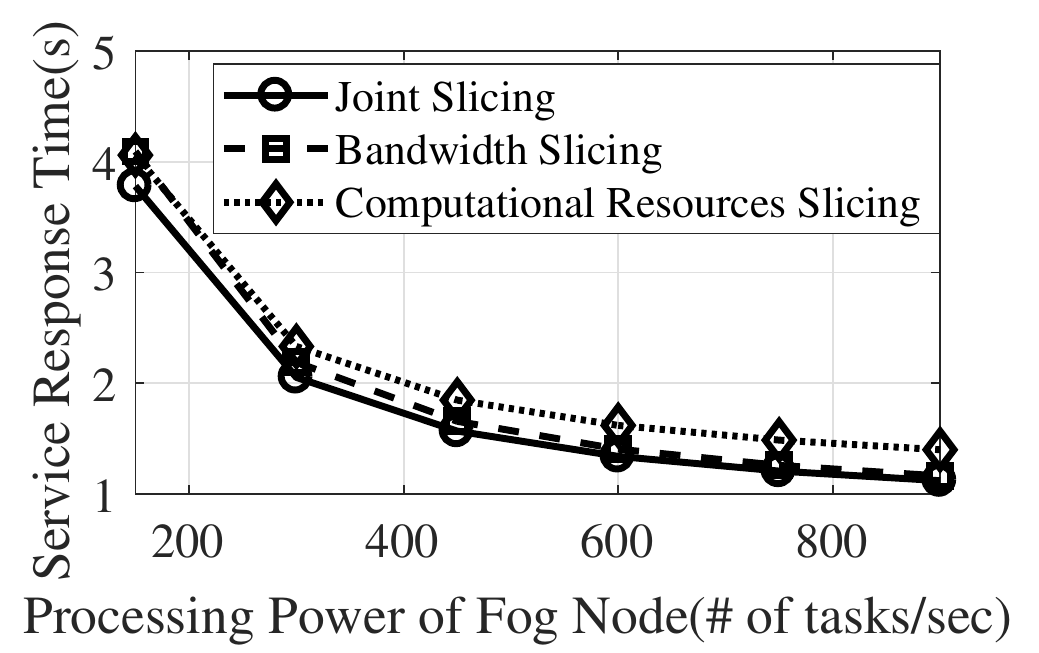}
 \vspace{-0.3in}
 \caption{Comparison of Three Architectures under Different Processing Rate of Each Fog Node of Area1}
 \label{fig5}
 \end{minipage}%

 }
 \subfigure{
 \begin{minipage}[t]{0.19\linewidth}
 \centering
 \includegraphics[width=3.8cm]{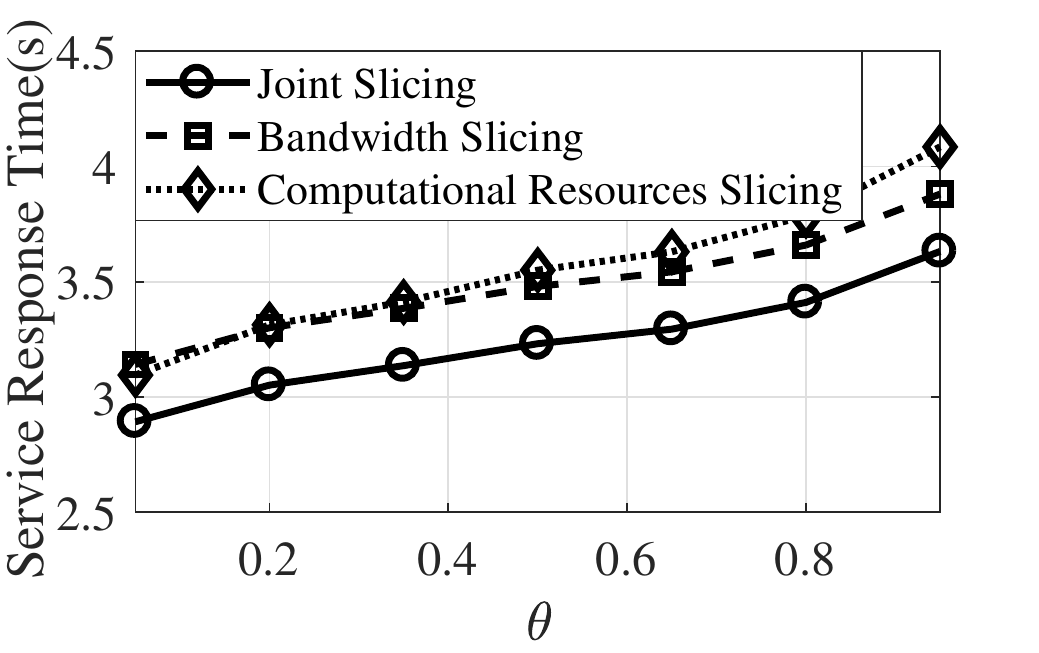}
 \vspace{-0.3in}
 \caption{Comparison of Three Architectures under Different $\theta$ of Area1}
 \label{fig6}
 \end{minipage}%
 }
 \vspace{-0.33in}
 \end{figure*}

%

We first evaluate the convergence performance of Algorithm 1. In Figure \ref{fig3}, we compare the interior-point algorithm with our proposed algorithm 1 (DistADMM-PVS) under different number of iterations. The interior-point algorithm has been widely applied in communication network systems \cite{8382242}. We can observe that our proposed Algorithm 1 can converge to a close neighborhood of the minimum latency within the first few iterations. It can offer much faster convergence performance compared to the interior-point algorithm.

%
%
%
%

We compare three architectures in different kinds of areas, as shown in Figure \ref{fig_area_comparison}. We can observe that our proposed joint slicing architecture performs better than other network slicing architecture in all three kinds of areas, which proves that our architecture has significant regional applicability, and can apply to different areas. We have the same characteristics for three architectures in all areas, so we only cover simulation results of area 1 in detail, as shown in Figure \ref{fig4}, Figure \ref{fig5} and Figure \ref{fig6}.

In Figure \ref{fig4}, we fix the processing power reserved for each fog node and the value of $\theta$ to evaluate the service response time under different bandwidth reserved for each BS. We can observe that the service response time decreases with the total reserved bandwidth. We can also observe that when the bandwidth of BSs is limited, the bandwidth slicing offers a better performance than computational resource slicing. However, as the bandwidth of each BS increases, the service response time of computational resource slicing starts to decrease much faster than that of the bandwidth slicing. This is because in our simulation, we fix the computational resource. In this case, when each BS has limited bandwidth, the communication latency dominates the overall latency. Therefore, applying bandwidth slicing to reduce the communication latency can have a higher impact than optimizing the computational resources in fog nodes for reducing the overall service response time. When the bandwidth of each BS becomes sufficient, the queuing delay will dominate overall latency. In this case, the computational resource slicing will become more useful to reduce the service response time.

In Figure \ref{fig5}, we fix the bandwidth reserved for each BS and value of $\theta$ to compare the service response time of three network slicing scenarios under different processing power reserved for each fog node. We can observe that the service response time decreases with the processing power reserved for each fog node. Similarly, we observe that when the computational resource of fog nodes is limited, the computational resource slicing offers a better performance than bandwidth slicing. However, as the computational resource of each fog node becomes sufficient, bandwidth slicing starts to decrease faster than the computational resource slicing. This is because in Figure \ref{fig5}, the bandwidth has been fixed. When each fog node has limited processing power, the queuing delay dominates the overall latency. When the processing power of each fog node increases, the communication latency starts to dominate the overall latency.

Similarly, in Figure \ref{fig6}, we fix the processing power reserved for each fog node and the bandwidth reserved for each BS to compare the service response time under different values of $\theta$. We observe that the service response time increases with $\theta$. This is because when $\theta$ increases, the total number of task units that need to be transported by BSs becomes larger. This will cause a higher communication delay, resulting in a higher service response time. We also observe that as $\theta$ increases, the service response time of bandwidth slicing starts to increase much slower than that of the computational resource slicing. This is because we fix both processing power reserved for each fog node and the bandwidth reserved for each BS. In this case, the number of task units for each service from each BS increases with $\theta$. When $\theta$ is small, bandwidth allocated to each task unit is sufficient and the queuing delay dominates the overall latency. When $\theta$ becomes larger, bandwidth allocated to each task unit is limited and the communication latency will start to dominate the overall latency.
\vspace{-0.12in}
\section{Conclusion}
\label{Section_Conclusion}
In this paper, we have investigated the distributed network slicing for a 5G system consisting of a set of BSs offering wireless communication services coexisting with a fog computing network performing computationally-intensive tasks. We propose a novel distributed framework based on a new control plane entity, regional orchestrator (RO), which can be deployed between BSs and fog nodes to coordinate and control their bandwidth and computational resources. We have also proposed a distributed resource allocation algorithm to distributedly coordinate the resource allocation of bandwidth of BSs and computational resources of fog nodes. Our simulation results show that the proposed algorithm can approach the global optimal solution with fast convergence rate. Moreover, our proposed distributed network slicing architecture can significantly improve network latency performance.

\section*{Acknowledgment}
The authors would like to acknowledge the support from National Key R\&D Program of China (2017YFE0121600).

\vspace{-0.15in}
\bibliography{reference}
\bibliographystyle{IEEEtran}

\end{document}